\documentclass[aps,preprint]{revtex4}

\usepackage{graphics,graphicx}
\usepackage{amsmath,amssymb}
\usepackage{epsfig}
\usepackage{txfonts}
\usepackage{subfigure}
\usepackage{xcolor}
\usepackage{epstopdf}
\usepackage{hyperref}

\begin{document}

\title{Linear Stability of \texorpdfstring{$f(R,\phi,X)$}{f(R,phi,X)} Thick Branes: Tensor Perturbations}
\author{Zheng-Quan Cui$^{1,2,}$\footnote{Email: cuizhq2017@lzu.edu.cn}}
\author{Yu-Xiao Liu$^{1,2,3,}$\footnote{Email: liuyx@lzu.edu.cn}}
\author{Bao-Min Gu$^{1,2,}$\footnote{Email: gubm15@lzu.edu.cn}}
\author{Li Zhao$^{1,2,}$\footnote{Email: lizhao@lzu.edu.cn, corresponding author}}

\affiliation{\small{$^{1}$ Institute of Theoretical Physics, Lanzhou University,\\
                           222 South Tianshui Road, Lanzhou 730000, China\\
                    $^{2}$ Research Center of Gravitation, Lanzhou University, \\
                           222 South Tianshui Road, Lanzhou 730000, China\\
                    $^{3}$ Key Laboratory for Magnetism and Magnetic Materials of the Ministry of Education, Lanzhou University,\\
                           222 South Tianshui Road, Lanzhou 730000, China}}

\begin{abstract}
We explore thick branes in $f(R,\phi,X)$ gravity. We obtain the linear tensor perturbation equation of $f(R,\phi,X)$ branes and show that the branes are stable against the tensor perturbations under the condition of $\frac{\partial f(R,\phi,X)}{\partial R}>0$. In order to obtain thick brane solutions of the fourth-order field equations in this theory, we employ the reconstruction technique. We get exact solutions of the specific $f(R,\phi,X)$ thick brane generated by a non-canonical scalar field. It is shown that the zero mode of the graviton for the thick brane is localized under certain conditions. This implies that the four-dimensional Newtonian potential is recovered on the brane. The effects of the Kaluza-Klein modes of the graviton for the $f(R,\phi,X)$ thick brane are also discussed.
\end{abstract}


\maketitle

\section{Introduction}

In extra-dimensional scenarios, some developments with large extra dimensions have gained considerable attention. Domain wall model~\cite{Rubakov:1983rsd} and Randall-Sundrum (RS) II model~\cite{Randall:1999rsa} are two typical examples of them. Rubakov and Shaposhnikov showed that fermions can be confined inside a domain wall generated by a scalar field in the absence of gravitation~\cite{Rubakov:1983rsd}. While Randall and Sundrum proposed that four-dimensional matter fields are confined on a thin brane embedded in five-dimensional anti-de Sitter (AdS) space-time~\cite{Randall:1999rsa}. It was also shown that the zero mode of the graviton is localized on the brane and contributes to the Newtonian potential, while the Kaluza-Klein (KK) modes of the graviton lead to corrections to the Newtonian potential. However, in the original RS II model, the brane is singular as the bulk curvature scalar is divergent at its location, and the extrinsic curvature satisfies the Israel junction conditions on the brane. Such conditions are generally not applicable or feasible for some higher-derivative theories of gravity~\cite{Chen:2012clzww}. These considerations have stimulated the development of thick brane models by considering the thickness of brane (for early works see Refs.~\cite{Goldberger:1999gw,Gremm:2000gf,
Gremm:2000gt,DeWolfe:2000dfgk,Csaki:2000cehs,Kobayashi:2002kks,Giovannini:2001gg,Giovannini:2002g,Giovannini:2003g}). In fact, the thick brane scenario is a natural generalization of domain wall model and RS II model. For reviews of topics related to the thick brane scenario, see Refs.~\cite{Dzhunushaliev:2010dfm,Liu:2017l}.

Some thick brane models in general relativity (GR) have been discussed, and trapping of various matter fields is realized on the brane. Although GR is widely accepted as a fundamental theory to describe the geometric properties of space-time, there are early attempts beginning in the 1920's by Weyl and Eddington who started to consider higher-order modifications of GR. Since then, modified theories of gravity have been explored in different ways (for reviews see~\cite{Nojiri:2007no,Sotiriou:2010sf,De Felice:2010dt}). In higher-order theories of gravitation, $f(R)$ theory which has achieved great successes was applied to the thick brane scenario~\cite{Afonso:2007abmp,
Dzhunushaliev:2010dfkk,Zhong:2011zly,Liu:2011lzzl,Liu:2012llw,Bazeia:2014blmps,Xu:2015xzyl,Bazeia:2015blm,
Bazeia:2015bllmo,Yu:2016yzgl,Gu:2017gzyl,Zhong:2016zl,Zhong:2017zl}. The thick brane scenario has also been explored from other perspectives, such as scalar-tensor theory, $K$-field (a non-canonical scalar field) theory~\cite{Bogdanos:2006bdt,Herrera-Aguilar:2012hmmq,Liu:2012lcgz,Guo:2012glzc,Yang:2012ylzdw,Ahmed:2013ag,
German:2014ghmqd,Bazeia:2007blmo,Adam:2008agsw,Adam:2008agksw,Bazeia:2009bglm,Zhong:2013zl,Zhong:2014zlzn,Zhong:2014zl}. $K$-field theory have offered an alternative mechanism for early time inflation~\cite{Armendariz-Picon:1999adm,
Garriga:1999gm,Chiba:2000coy,Armendariz-Picon:2001ams}, and were also discussed in topological defects. Whether the Lagrangian of the non-canonical scalar field yields to other extensions considering gravitation or not is worth to verify. A natural treatment is $f(R,\phi,X)$ gravity~\cite{De Felice:2010dt,Tsujikawa:2007t,Bahamonde:2015bbls} which abandons any assumptions on the theory of gravitation with the exception of being second-order and includes $f(R)$ gravity, general scalar-tensor gravity, and non-canonical scalar field theories. We expect to investigate thick branes in $f(R,\phi,X)$ gravity. Such a theory contains the departures from minimal coupling theories, canonical scalar field theories, and second-order derivative theories, etc. It would be interesting to study the effects on braneworld models, both on the background structure and the perturbations.

It is well known that the configuration and stability of thick branes are two significant issues. Firstly, brane systems should be stable at least under linear perturbations. Furthermore, to coincide with gravitational experiments, the four-dimensional Newtonian potential should be recovered, which indicates that the zero mode of the graviton is localized on the brane. To this end, we investigate the stability of tensor perturbations and the localization of the graviton. With regard to thick brane configurations, they are usually suggested as topological defects, such as domain walls (kink-like configurations). It is known that kinks or domain walls are the simplest solitons and hence they are invaluable for learning about non-perturbative aspects of field theories~\cite{Vachaspati:2010v}. Soliton solutions have been found for thick branes generated by one or more scalar fields, see Refs.~\cite{Gremm:2000gf,Gremm:2000gt,Kobayashi:2002kks,Dzhunushaliev:2010dfkk,Afonso:2007abmp,Liu:2011lzzl,
Bazeia:2014blmps,Bazeia:2015blm,Bazeia:2015bllmo,Bogdanos:2006bdt,Herrera-Aguilar:2012hmmq,German:2014ghmqd,
Liu:2012lcgz,Guo:2012glzc,Ahmed:2013ag,Bazeia:2007blmo,Adam:2008agsw,Bazeia:2009bglm,Zhong:2014zl,Peyravi:2016prl} or reviews~\cite{Dzhunushaliev:2010dfm,Liu:2017l} and references therein. However, as the equations of motion of solitons and of gravitational field are generally non-linear and even higher-order in $f(R,\phi,X)$ gravity, some effective approaches necessarily are proposed to obtain analytical brane solutions. Reconstruction techniques which are extensively employed in cosmology are effective approaches for this purpose. For instance, a reconstruction technique has been applied to explore a general Friedmann-Lema\^{\i}tre-Robertson-Walker domain wall universe~\cite{Higuchi:2014hn}. In this paper, we seek for the original actions starting from the configurations of thick branes in $f(R,\phi,X)$ gravity by means of another reconstruction technique. We suppose that the scalar field has a domain wall configuration and the warp factor is a smooth function. As we will demonstrate, thick brane configuration with a non-canonical scalar field is supported in $f(R,\phi,X)$ gravity.

This paper is organized as follows. In section~\ref{sec:model}, we consider a general action in five-dimensional space-time and give the general equations of motion in the flat thick brane scenario. In section~\ref{sec:perturbation}, we show that $f(R,\phi,X)$ thick branes are stable against the tensor perturbations. In section~\ref{sec:brane}, we investigate the $f(R,\phi,X)$ thick brane with a domain wall configuration. In section~\ref{sec:localization}, we analyze the localization of the zero mode of the graviton for a specific form of $f(R,\phi,X)$. In section~\ref{sec:potential}, the effects of the massive KK modes of the graviton to the Newtonian potential is discussed for the $f(R,\phi,X)$ thick brane with a domain wall configuration. Finally, discussion and conclusions are given in section~\ref{sec:conclusions}.

\section{Action and field equations}
\label{sec:model}

Consider the following five-dimensional action within the context of the generalized modified theories of gravity, the $f(R,\phi,X)$ gravity~\cite{De Felice:2010dt,Tsujikawa:2007t,Bahamonde:2015bbls},
\begin{equation}
S=\int\mathrm{d}^5x\sqrt{-g}\frac{1}{2\kappa_5^2}f(R,\phi,X),\label{action}
\end{equation}
where $\kappa_5$ is the five-dimensional coupling constant, and $f(R,\phi,X)$ is an arbitrary function of the curvature scalar $R$, scalar field $\phi$, and kinetic term $X=-\frac{1}{2}g^{MN}\nabla_M\phi\nabla_N\phi$~\footnote{We are using the conventions $R^P_{MQN}=\partial_Q\Gamma^P_{MN}-\partial_N\Gamma^P_{MQ}+\Gamma^P_{QL}\Gamma^L_{MN}-\Gamma^P_{NL}\Gamma^L_{MQ}$ and $R_{MN}=R^L_{MLN}$. The metric signature is $(-,+,+,+,+)$.}. Throughout the paper, capital Latin letters $M,N,\ldots$ represent the five-dimensional coordinate indices running over $0,1,2,3,5$, and lower-case Greek letters $\mu,\nu,\ldots$ represent the four-dimensional coordinate indices running over $0,1,2,3$.

The variations of action~\eqref{action} with respect to the metric $g_{MN}$ and the scalar field $\phi$ respectively yield the following field equations
\begin{equation}\label{eq:fieldeq1}
f_{R}G_{MN}=\frac{1}{2}\left(f-Rf_{R}\right)g_{MN}+\nabla_M\nabla_Nf_{R}-g_{MN}\nabla_A\nabla^Af_{R}+
\frac{1}{2}f_{X}\nabla_M\phi\nabla_N\phi,
\end{equation}
\begin{equation}\label{eq:fieldeq2}
\nabla_M\left(f_{X}\nabla^M\phi\right)+f_{\phi}=0.
\end{equation}
Here $f_{R}=\partial f/\partial R$, $f_{X}=\partial f/\partial X$, and $f_{\phi}=\partial f/\partial\phi$.

In this paper, we are interested in static flat branes with four-dimensional Poincar\'{e} symmetry, for which the line element is given by
\begin{equation}\label{linee}
\mathrm{d}s^2=\mathrm{e}^{2A(y)}\eta_{\mu\nu}\mathrm{d}x^\mu\mathrm{d}x^\nu+\mathrm{d}y^2,
\end{equation}
where $\mathrm{e}^{2A(y)}$ is the warp factor, $\eta_{\mu\nu}$ is the four-dimensional Minkowski metric, and $y=x^5$ is the extra-dimensional coordinate. Note that the warp factor $\mathrm{e}^{2A(y)}$ and the background scalar field $\phi$ are merely functions of $y$ for static flat branes. Denoting $a(y)=\mathrm{e}^{A(y)}$ and using the metric ansatz~\eqref{linee}, Eqs.~\eqref{eq:fieldeq1} and \eqref{eq:fieldeq2} are reduced to
\begin{subequations}\label{eq:eq1MN}
\begin{align}
f+2\left(3a^{-2}a'^{2}+a^{-1}a''\right)f_{R}-6a^{-1}a'f_{R}'-2f_{R}''=0,\label{eq:eq1compuv}\\
f+8a^{-1}a''f_{R}-8a^{-1}a'f_{R}'+f_{X}\phi'^{2}=0,\label{eq:eq1comp55}
\end{align}
\end{subequations}
and
\begin{equation}\label{eq:eq2phi}
f_{X}'\phi'+\left(\phi''+4a^{-1}a'\phi'\right)f_{X}+f_{\phi}=0,
\end{equation}
respectively, where the prime denotes the derivative with respect to the extra-dimensional coordinate $y$. It is worth pointing out that, however, only two equations are independent in Eqs.~\eqref{eq:eq1MN} and \eqref{eq:eq2phi}. Before seeking solutions of these brane systems, we analyze the tensor perturbations in the following section for addressing the localization of the graviton.

\section{Tensor perturbations}
\label{sec:perturbation}

The perturbation to the metric~\eqref{linee} can be decomposed into three kinds of modes, namely the transverse-traceless (TT) tensor modes, transverse vector modes, and scalar modes. To linear order, the three kinds of modes are decoupled from each other, one can treat them separately~\cite{Giovannini:2001gg,Giovannini:2002g,Giovannini:2003g,Weinberg:2008w}. In the following, we investigate the linear stability of the TT tensor perturbations.

We consider the following tensor perturbations:
\begin{equation}\label{perturbation}
\mathrm{d}s^2=a^2(y)(\eta_{\mu\nu}+h_{\mu\nu})\mathrm{d}x^\mu\mathrm{d}x^\nu+\mathrm{d}y^2,
\end{equation}
or
\begin{equation}
g_{MN}=
\begin{bmatrix}\label{matrix}
a^2(y)(\eta_{\mu\nu}+h_{\mu\nu}) & 0 \\ 0 & 1 \\
\end{bmatrix},
\end{equation}
where $a^2(y)h_{\mu\nu}$ is the linear part of the tensor perturbations. From Eq.~\eqref{matrix}, one has
\begin{align}\label{deltag}
\delta g_{\mu\nu}=a(y)^2h_{\mu\nu},\quad \delta g_{\mu 5}=\delta g_{55}=0.
\end{align}
Here $\delta$ refers to the linear order perturbations. The inverse of the metric perturbation~\eqref{matrix} takes the form
\begin{equation}
g^{MN}=
\begin{bmatrix}\label{invmatrix}
a^{-2}(y)(\eta^{\mu\nu}-h^{\mu\nu}) & 0 \\ 0 & 1 \\
\end{bmatrix},
\end{equation}
where $h^{\mu\nu}=\eta^{\mu\rho}\eta^{\nu\sigma}h_{\rho\sigma}$. The linear tensor perturbations $h_{\mu\nu}$ depend on all the coordinates, i.e., $h_{\mu\nu}=h_{\mu\nu}(x^{\mu},y)$.

Taking Eqs.~\eqref{matrix} and \eqref{invmatrix} into account, the linear perturbations of the Ricci tensor and curvature scalar are obtained as
\begin{align}\label{perturbedRMNR}
\delta R_{\mu\nu}&=\frac{1}{2}\left(\partial_{\nu}\partial_{\sigma}h_{\mu}^{\sigma}+\partial_{\mu}\partial_{\sigma}h_{\nu}^{\sigma}-
\square^{(4)}h_{\mu\nu}-\partial_{\mu}\partial_{\nu}h\right) \nonumber \\
&~-\left(3a'^2+aa''\right)h_{\mu\nu}-2aa'h_{\mu\nu}'-\frac{1}{2}a^2h_{\mu\nu}''-\frac{1}{2}aa'\eta_{\mu\nu}h',\nonumber\\
\delta R_{\mu5}&=\frac{1}{2}\partial_y\left(\partial_\sigma h_\mu^\sigma-\partial_\mu h\right),\quad
\delta R_{55}=-\frac{1}{2}\left(2a^{-1}a'h'+h''\right),\nonumber\\
\delta R&=\delta\left(g^{MN}R_{MN}\right)=
a^{-2}\left(\partial_{\mu}\partial_{\nu}h^{\mu\nu}-\square^{(4)}h\right)-5a^{-1}a'h'-h'',
\end{align}
where $\square^{(4)}=\eta^{\mu\nu}\partial_{\mu}\partial_{\nu}$ is the four-dimensional d'Alembert operator, and $h=\eta^{\mu\nu}h_{\mu\nu}$.

Generally, the equation~\eqref{eq:fieldeq1} for the linear tensor perturbations is arrived at
\begin{align}\label{eq:perturbedfieldeq1}
\delta f_{R}G_{MN}+f_{R}\delta G_{MN}=
\frac{1}{2}\left[\left(\delta f-\delta Rf_{R}-R\delta f_{R}\right)g_{MN}+\left(f-Rf_{R}\right)\delta g_{MN}\right] \nonumber\\
+\delta\left(\nabla_M\nabla_Nf_{R}\right)-\delta\left(g_{MN}\nabla_A\nabla^Af_{R}\right)+
\frac{1}{2}\delta f_{X}\nabla_M\phi\nabla_N\phi.
\end{align}
Since $f=f(R,\phi,X)$ is a function of $R$, $\phi$, and $X$, the $f(R,\phi,X)$ for the linear tensor perturbations is derived as
\begin{equation}\label{deltaf}
\delta f(R,\phi,X)=\frac{\partial{f}}{\partial{R}}\delta R.
\end{equation}
It is easily found that $\delta f_{R}$, $\delta f_{\phi}$, and $\delta f_{X}$ are determined in the same way as Eq.~\eqref{deltaf}.

With the help of the expansions of $\nabla_M\nabla_Nf_{R}$ and $g_{MN}\nabla_A\nabla^Af_{R}$:
\begin{subequations}\label{relations}
\begin{align}
\nabla_M\nabla_Nf_{R}&=\left(\partial_M\partial_N-\Gamma^P_{NM}\partial_P\right)f_{R},\\
g_{MN}\square^{(5)}f_{R}&=g_{MN}\nabla_A\nabla^Af_{R}=g_{MN}g^{AB}\left(\nabla_A\nabla_Bf_{R}\right),
\end{align}
\end{subequations}
where $\square^{(5)}=g^{AB}\nabla_{A}\nabla_{B}$ is the five-dimensional d'Alembert operator, one writes the two terms $\delta\left(\nabla_M\nabla_Nf_{R}\right)$ and $\delta\left(g_{MN}\nabla_A\nabla^Af_{R}\right)$ in Eq.~\eqref{eq:perturbedfieldeq1} as
\begin{subequations}
\begin{align}
\delta\left(\nabla_M\nabla_Nf_{R}\right)&=
\left(\partial_M\partial_N-\Gamma^P_{NM}\partial_P\right)\delta f_{R}-\delta\Gamma^P_{NM}\partial_Pf_{R},\label{expanddelta1}\\
\delta\left(g_{MN}\nabla_A\nabla^Af_{R}\right)&=\delta\left(g_{MN}g^{AB}\nabla_A\nabla_Bf_{R}\right) \nonumber \\
&=\delta g_{MN}\square^{(5)}f_{R}+g_{MN}\delta g^{AB}\left(\nabla_A\nabla_Bf_{R}\right)+ g_{MN}g^{AB}\delta\left(\nabla_A\nabla_Bf_{R}\right).\label{expanddelta2}
\end{align}
\end{subequations}

According to the TT conditions
\begin{align}\label{TTcondition}
\partial_\mu h^\mu_{\nu}=0,\quad h=\eta^{\mu\nu}h_{\mu\nu}=0,
\end{align}
Eq.~\eqref{perturbedRMNR} is simplified as
\begin{align}\label{TTRMNR}
\delta R_{\mu\nu}&=-\frac{1}{2}\square^{(4)}h_{\mu\nu}-\left(3a'^2+aa''\right)h_{\mu\nu}-2aa'h_{\mu\nu}'
-\frac{1}{2}a^2h_{\mu\nu}'',\nonumber \\
\delta R_{\mu5}&=0,\quad \delta R_{55}=0,\quad \delta R=0.
\end{align}
With the above result $\delta R=0$, one has
\begin{align}\label{TTdeltaffRfX}
\delta f(R,\phi,X)=0,\quad
\delta f_{R}(R,\phi,X)=0,\quad
\delta f_{X}(R,\phi,X)=0.
\end{align}
Then, the $\mu\nu$-components of $\delta\left(\nabla_M\nabla_Nf_{R}\right)$, $\delta\left(g_{MN}\nabla_A\nabla^Af_{R}\right)$, and $\delta G_{MN}$ can be calculated as
\begin{subequations}\label{TTmunu}
\begin{align}
\delta\left(\nabla_\mu\nabla_\nu f_{R}\right)&=
f_{R}'\left(aa'h_{\mu\nu}+\frac{1}{2}a^2h_{\mu\nu}'\right),\label{TTexpanddelta1} \\
\delta\left(g_{\mu\nu}\nabla_A\nabla^Af_{R}\right)&=
a^2\left(4a^{-1}a'f_{R}'+f_{R}''\right)h_{\mu\nu},\label{TTexpanddelta2} \\
\delta G_{\mu\nu}&=-\frac{1}{2}\square^{(4)}h_{\mu\nu}+3\left(a'^{2}+aa''\right)h_{\mu\nu}-2aa'h_{\mu\nu}'-
\frac{1}{2}a^2h_{\mu\nu}''.\label{TTG}
\end{align}
\end{subequations}

Substituting Eqs.~\eqref{TTdeltaffRfX} and \eqref{TTmunu} into Eq.~\eqref{eq:perturbedfieldeq1}, one gets the $\mu\nu$-components of the perturbed field equation:
\begin{align}\label{perturbedequv}
\frac{1}{2}a^2\left[f+2\left(3a^{-2}a'^{2}+a^{-1}a''\right)f_{R}-6a^{-1}a'f_{R}'-2f_{R}''\right]h_{\mu\nu} \nonumber \\
+f_{R}\left(\frac{1}{2}\square^{(4)}h_{\mu\nu}+2aa'h_{\mu\nu}'+\frac{1}{2}a^2h_{\mu\nu}''\right)+
\frac{1}{2}a^2f_{R}'h_{\mu\nu}'=0.
\end{align}
Thus, we eventually arrive at the main equation for the tensor perturbations by noticing Eq.~\eqref{eq:eq1compuv}:
\begin{equation}\label{eq:perturbedequvend}
f_{R}\left(\square^{(4)}h_{\mu\nu}+4a a'h_{\mu\nu}'+a^{2}h_{\mu\nu}''\right)+ a^{2}f_{R}'h_{\mu\nu}'=0,
\end{equation}
which can also be written as
\begin{equation}\label{eq:perturbedeqsim}
\square^{(5)}h_{\mu\nu}=f_{R}^{-1}f_{R}'\partial_{y}h_{\mu\nu}.
\end{equation}

With the coordinate transformation $\mathrm{d}z=a^{-1}\mathrm{d}y$ between the coordinate $z$ and $y$, Eq.~\eqref{eq:perturbedequvend} turns into
\begin{equation}
\left[\partial_z^{2}+\left(3a^{-1}\partial_za+f_{R}^{-1}\partial_zf_{R}\right)\partial_z+
\square^{(4)}\right]h_{\mu\nu}=0.
\end{equation}
Next, we perform the KK decomposition $h_{\mu\nu}(x^\rho,z)=\epsilon_{\mu\nu}(x^{\rho})\psi(z)f(z)$ with $f(z)=a^{-3/2}(z)f^{-1/2}_{R}$, which requires $f_{R}>0$. Then we obtain the Klein-Gordon equation for the four-dimensional part $\epsilon_{\mu\nu}(x^\rho)$:
\begin{equation}\label{eq:KG}
\left(\square^{(4)}+m^2\right)\epsilon_{\mu\nu}(x^{\rho})=0,
\end{equation}
and the Schr\"{o}dinger-like equation for the extra-dimensional part $\psi(z)$:
\begin{equation}\label{eq:Schrodinger}
\left[-\partial_z^2+W(z)\right]\psi(z)=m^2\psi(z).
\end{equation}
Here the effective potential $W(z)$ has the following form
\begin{align}
W(z)&=\frac34\frac{\left(\partial_{z}a\right)^2}{a^2}+\frac32\frac{\partial_{z}\partial_{z}a}{a}+
\frac32\frac{\partial_{z}a}{a}\frac{\partial_{z}f_{R}}{f_{R}}+\frac12\frac{\partial_{z}\partial_{z}f_{R}}{f_{R}}-
\frac14\frac{\left(\partial_{z}f_{R}\right)^2}{f_{R}^2}\label{epotential} \\
&=\Omega^2+\partial_z\Omega,\label{vepotential}
\end{align}
with
\begin{equation}\label{Omega}
\Omega=\frac32\frac{\partial_za}{a}+\frac12\frac{\partial_zf_R}{f_R}.
\end{equation}
The equation~\eqref{eq:Schrodinger} can be factorized as $\Theta\Theta^{\dagger}\psi(z)=m^2\psi(z)$ with
\begin{equation}
\Theta=\partial_z+\Omega,\quad \Theta^{\dagger}=-\partial_z+\Omega.
\end{equation}
Hence, the factorization of the Schr\"{o}dinger-like equation ensures that there is no the gravitational tachyon mode with $m^2<0$. In other words, the system is stable against the tensor perturbations~\eqref{perturbation} under the condition of $f_{R}>0$. It should be addressed that $f$ is an arbitrary function of $R$, $\phi$, and $X$, such as, $f=q(R)+X-V(\phi)$~\cite{Afonso:2007abmp}, $f=p(\phi)R+X-V(\phi)$~\cite{Bogdanos:2006bdt}, $f=R+L(\phi,X)$~\cite{Bazeia:2009bglm}, which are respectively corresponding to the $f(R)$ thick branes, scalar-tensor thick branes, and $K$-field thick branes generated by the background scalar field $\phi$.

The linear tensor perturbations are stable as previous works~\cite{Gu:2017gzyl, Zhong:2011zly}, while our case includes the non-minimal coupling between the non-canonical scalar field and gravitation. Actually, the equation of the tensor modes remains second-order, whereas the equation for the scalar modes is fourth-order. From this point of view, there is no modification to the tensor part. However, as we will show in sections~\ref{sec:localization} and~\ref{sec:potential}, the behavior of gravitons is definitely modified because of the background structure. This can be seen from Eqs.~\eqref{eq:Schrodinger} and~\eqref{epotential}, which contain the contribution from the background quantities $\phi$ and $X$.

\section{\texorpdfstring{$f(R,\phi,X)$}{f(R,phi,X)} thick brane solutions}
\label{sec:brane}

In $(1+1)$-dimensional field theory, a canonical scalar field has static kink solutions~\cite{Dashen:1974dhn}. According to Derrick's theorem~\cite{Derrick:1964d}, if the number of spatial dimensions is larger than one, there are no static solutions of a canonical scalar field with finite energy. For the sake of obtaining domain wall solutions, three possibilities were considered in the works of other authors:
\begin{itemize}
\item static solutions of a non-canonical scalar field with finite energy~\cite{Diaz-Alonso:1983da};
\item non-static solutions of a canonical scalar field with finite energy~\cite{Friedberg:1976fls};
\item static domain wall solutions of a canonical scalar field with infinite energy~\cite{Vachaspati:2010v}.
\end{itemize}
It should be mentioned that Rubakov and Shaposhnikov have found a static domain wall solution of the $\phi^4$ scalar field model whose kinetic term is canonical in five-dimensional flat space-time~\cite{Rubakov:1983rsd}. The domain wall solution of the scalar field takes the kink-like form $\phi(y)=v\tanh(ky)$, which is only a function of the extra-dimensional coordinate $y$. Nevertheless, it suffers from a drawback that gravitation is not considered in the theory. The $\phi^4$ model of a canonical scalar field in GR have not yet found an analytic domain wall solution. Furthermore, the containing of curvature corrections, such as $f(R)=R+\beta R^2$ gravity~\cite{Liu:2011lzzl}, Euler-Gauss-Bonnet combination~\cite{Corradini:2000ck,Giovannini:2001gt}, leads to analytic smooth solutions. Hence, the adding of higher curvature terms is an effective way to explore domain walls.

In this section, we investigate whether the generalized $f(R,\phi,X)$ gravity theories support domain wall configurations. We firstly postulate an abstract form of $f(R,\phi,X)$ in action~\eqref{action} from theoretical consideration. Essentially, general coordinate covariance does not forbid including non-minimal coupling invariant terms which are vanishing in flat space-time. Such terms in the action describe the non-minimal coupling between the scalar field and gravitation~\cite{Buchbinder:1992bos}. Furthermore, we suppose that the non-canonical scalar field generates the thick brane. One typical model under such constraints may be written in the following form
\begin{equation}\label{gphiR}
f(R,\phi,X)=p(\phi)q(R)+K(X)-V(\phi),
\end{equation}
where $K(X)$ is an arbitrary function of the canonical kinetic term $X$ and $V(\phi)$ is the potential of the scalar field $\phi$. In general, regarding $p(\phi)$~\cite{Bogdanos:2006bdt} and $q(R)$~\cite{Sotiriou:2010sf} as a series expansion, namely
\begin{subequations}\label{pphiqR}
\begin{align}
p(\phi)&=p(0)+\frac{1}{2}p''(0)\phi^2+\cdots,\label{pphi} \\
q(R)&=\cdots+\frac{\gamma_2}{R^2}+\frac{\gamma_1}{R}-2\Lambda+R+\frac{R^2}{\beta_2}+\frac{R^3}{\beta_3}+\cdots,\label{qR}
\end{align}
\end{subequations}
where the coefficients $\gamma_i$ and $\beta_i$ have appropriate dimension, one may find that the action contains a number of phenomenologically interesting terms. In what follows, for simplicity, we restrict the general functions $p(\phi)$ and $q(R)$ to be a quadratic function in $\phi$ and the linear and quadratic terms of $R$, respectively. Therefore, a typical form of $f(R,\phi,X)$ in action~\eqref{action} is given by
\begin{equation}\label{phi2R2}
f(R,\phi,X)=(1+\alpha\phi^2)(R+\beta R^2)+K(X)-V(\phi),
\end{equation}
where $\alpha$ and $\beta$ are arbitrary parameters.

\begin{figure}[htb]
\centering
\subfigure[\ $a(\tilde y)$]{\label{subfig:ay}
\includegraphics[width=2.6in]{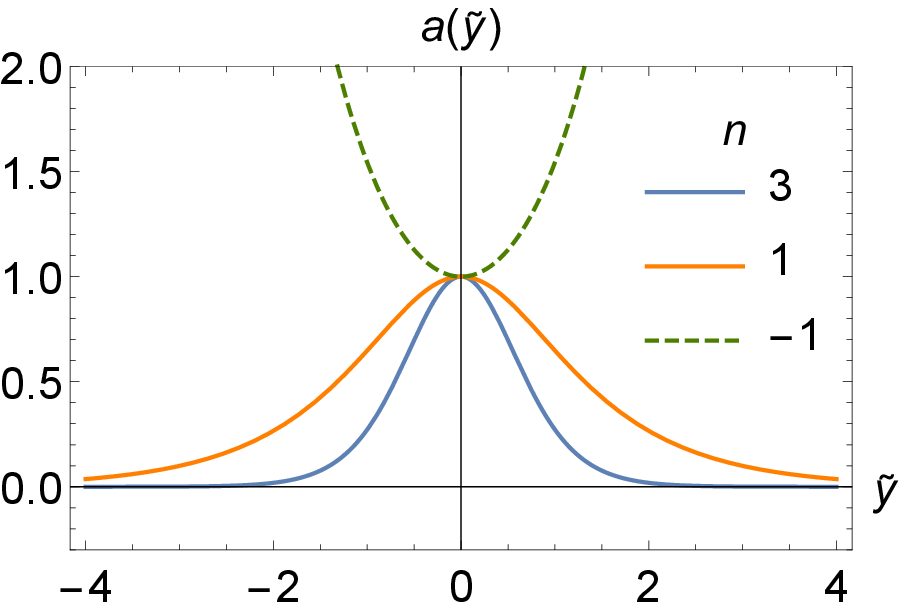}}
\hspace{0.5in}
\subfigure[\ $\phi(\tilde y)/v$]{\label{subfig:phiy}
\includegraphics[width=2.6in]{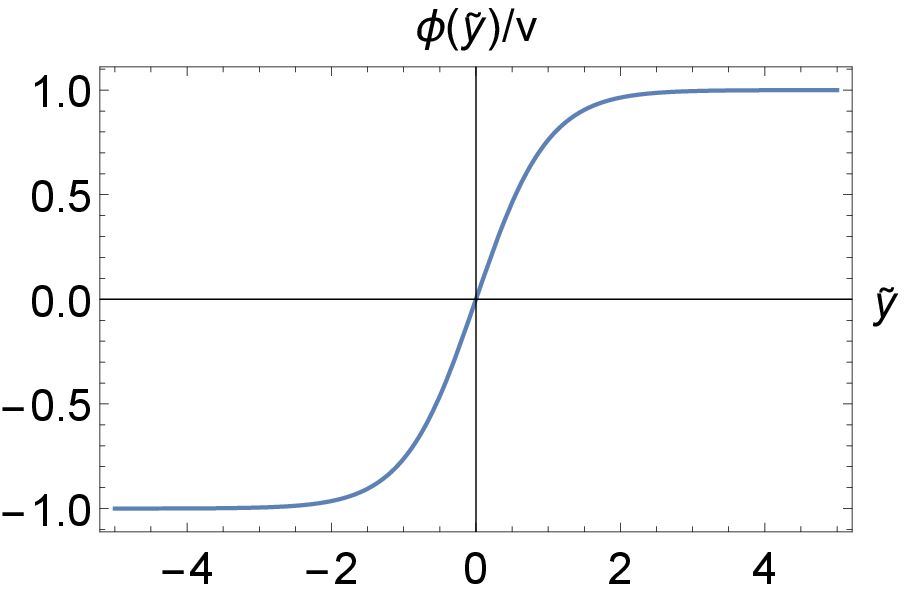}}
\caption{Plots of $a(\tilde y)$ and $\phi(\tilde y)/v$, where $a(\tilde y)$ is the warp factor in Eq.~\eqref{warpfactor} and $\phi(\tilde y)$ is the scalar field in Eq.~\eqref{kink}. The parameter is set to $n=(3,1,-1)$.}\label{fig:ay&phiy}
\end{figure}

By substituting the form of $f(R,\phi,X)$ in~\eqref{phi2R2} into Eqs.~\eqref{eq:eq1MN} and \eqref{eq:eq2phi}, we can obtain the explicit forms of the field equations. It is clear that there are four unknown quantities but only two independent field equations. Therefore, we give two of them and solve the other quantities. One usually gives the forms of $K(X)$ and $V(\phi)$. However, the equation~\eqref{eq:eq1compuv} for the warp factor $a(y)$ is fourth-order and hence is hard to be solved. In order to find out analytic brane solutions, we give the following $a(y)$ and $\phi(y)$ and solve $K(X)$ and $V(\phi)$:
\begin{align}
a(y)&=\cosh^{-n}{(ky)},\label{warpfactor} \\
\phi(y)&=v\tanh(ky).\label{kink}
\end{align}
Note that the above warp factor indicates that the bulk space-time is asymptotically AdS, which is essential for the localization of gravitation. The kink-like configuration~\eqref{kink} is a typical form of domain walls. The warp factor and scalar field are depicted in Fig.~\ref{fig:ay&phiy}. In order to display the trend of $a(y)$ and $\phi(y)$, we introduce the dimensionless quantities $\tilde y=ky$, $a(\tilde y)=\cosh^{-n}{(\tilde y)}$, and $\phi(\tilde y)/v=\tanh(\tilde y)$ in Fig.~\ref{fig:ay&phiy}. It is clear that the warp factor diverges if $n<0$ and converges if $n>0$, and the scalar field is a configuration of kink.

\begin{figure}[htb]
\centering
\subfigure[\ $V(\phi)$]{\label{subfig:Vphi:a}
\includegraphics[width=2.5in]{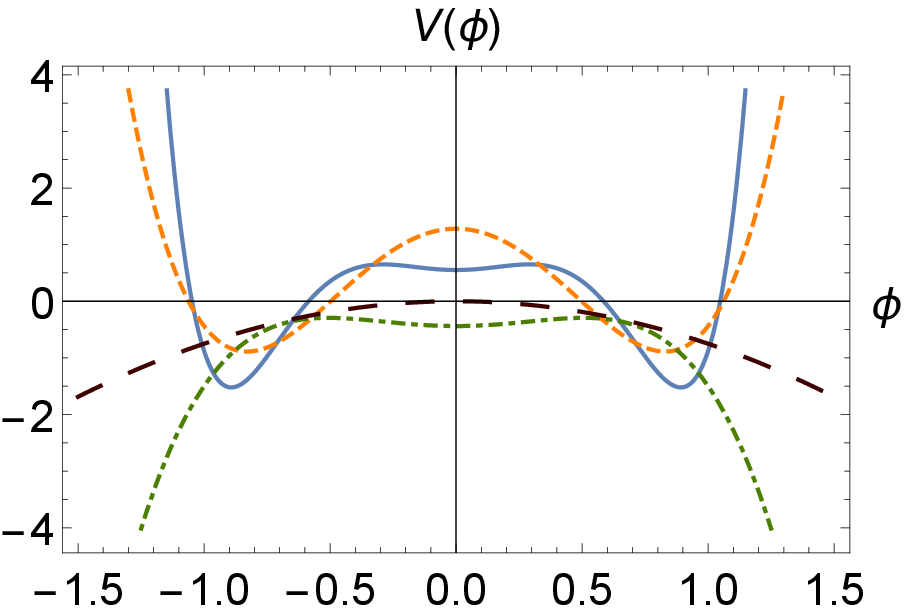}}
\hspace{0.1in}
\subfigure[\ Legend]{\label{subfig:l2}
\includegraphics[width=0.8in]{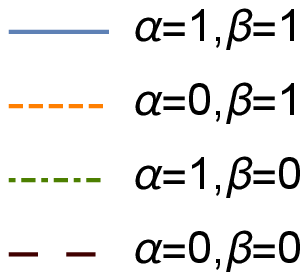}}
\hspace{0.1in}
\subfigure[\ $V(y)$]{\label{subfig:Vphi:a'}
\includegraphics[width=2.5in]{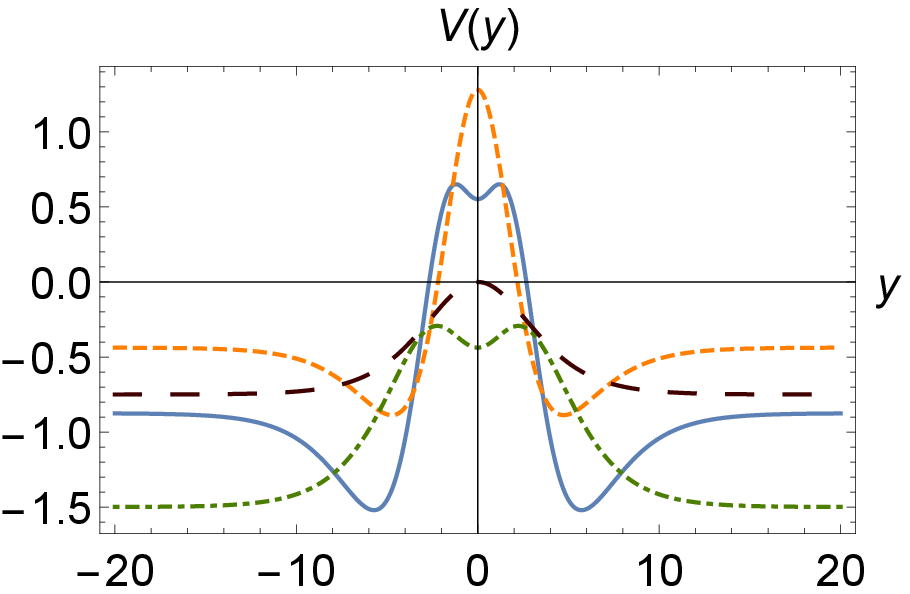}}
\caption{Plots of $V(\phi)$ and $V(y)$ in Eq.~\eqref{phi2R2Vphis} for $f(R,\phi,X)=(1+\alpha\phi^2)(R+\beta R^2)+K(X)-V(\phi)$. The legend of the parameters $\alpha$ and $\beta$ is given in Fig.~\ref{subfig:l2}. The parameters are set to $k=0.25$, $v=1$, $c=0$, and $n=1$.}\label{fig:Vphi&Vy}
\end{figure}

The substitution of Eqs.~\eqref{phi2R2}, \eqref{warpfactor}, and \eqref{kink} into Eqs.~\eqref{eq:eq1MN} gives the form of $K(X)$ and $V(\phi)$ (see Appendix for details)
\begin{subequations}\label{phi2R2_KV}
\begin{align}
K(X)&=A_X\sqrt{-X}+B_X\left(\sqrt{-X}\right)^2+C_X\left(\sqrt{-X}\right)^3+D_X,\label{phi2R2KXs} \\
V(\phi)&=A_{\phi}\phi^2+B_{\phi}\phi^4+C_{\phi}\phi^6+D_{\phi}.\label{phi2R2Vphis}
\end{align}
\end{subequations}
The potential $V$ in Eq.~\eqref{phi2R2Vphis} is shown in Fig.~\ref{fig:Vphi&Vy} in both the $\phi$ and $y$ coordinates.
One can also obtain a simple form of $f(R,\phi,X)$ by rescaling the scalar field. Note that the scalar field considered here is non-canonical.

For demonstrating the validity of the reconstruction technique, we illustrate whether the action of the domain wall in five-dimensional flat space-time is recovered or not. Taking the limit $n\to0$ which implies that the five-dimensional space-time tends to be flat, one will find the following result ($c=0$)
\begin{subequations}\label{n=0}
\begin{align}
  K(X)&=-4\alpha\left[2\sqrt{2}kv\sqrt{-X}-3\left(\sqrt{-X}\right)^2\right], \\
  V(\phi)&=-\frac{6\alpha k^2}{v^2} \left(\phi^2-v^2\right)^2.
\end{align}
\end{subequations}
Therefore, the assumption~\eqref{phi2R2} would reduce to $f(R,\phi,X)=K(X)-V(\phi)$ with $K(X)$ and $V(\phi)$ given by~\eqref{n=0}. It is demonstrated that the domain wall solution~\eqref{kink} exists for the $\phi^4$ scalar field model in five-dimensional flat space-time.

In what follows, we consider the non-canonical scalar field interacting with gravitation. Note that for the special case of $\alpha=0$, $\beta=0$, and $c=0$, the solution~\eqref{phi2R2_KV} is simplified as
\begin{subequations}\label{solutionAlpha=0Beta=0}
\begin{align}
K(X)&=-\frac{6\sqrt{2}kn}{v}\sqrt{-X},\label{phi2R2KXss}\\
V(\phi)&=-\frac{12k^2n^2}{v^2}\phi^2,\label{phi2R2Vphiss}
\end{align}
\end{subequations}
and $f(R,\phi,X)$ reduces to
\begin{equation}
f(R,\phi,X)=R+K(X)-V(\phi),
\end{equation}
which is just the case of GR with the non-canonical scalar field. It is demonstrated, in the GR case, that the domain wall solution~\eqref{kink} can not be sustained with the non-canonical kinetic term~\eqref{phi2R2KXss} and the scalar field potential~\eqref{phi2R2Vphiss}.

Next, we consider two other kinds of $f(R,\phi,X)$ with $\alpha=0,~\beta\neq0$ and $\alpha\neq0,~\beta=0$. The former leads to the $f(R)$ gravity with $f(R,\phi,X)=(R+\beta R^2)+K(X)-V(\phi)$, and the latter is the scalar-tensor gravity with $f(R,\phi,X)=(1+\alpha\phi^2)R+K(X)-V(\phi)$. In the case of $\alpha=0$, we have
\begin{subequations}
\begin{align}
K(X)&=A_{X\beta}\sqrt{-X}+B_{X\beta}\left(\sqrt{-X}\right)^2+C_{X\beta},\label{phi0R2KXs}\\
V(\phi)&=A_{\phi\beta}\phi^2+B_{\phi\beta}\phi^4+C_{\phi\beta},\label{phi0R2Vphis}
\end{align}
\end{subequations}
where the coefficients $A_{X\beta}$, $B_{X\beta}$, $C_{X\beta}$, $A_{\phi\beta}$, $B_{\phi\beta}$, and $C_{\phi\beta}$ are related to the parameters $\beta$, $k$, $v$, and $c$, but are clearly irrelevant to $\alpha$.
For the second case, $f(R,\phi,X)=(1+\alpha\phi^2)R+K(X)-V(\phi)$, the solution reads as
\begin{subequations}
\begin{align}
K(X)&=A_{X\alpha}\sqrt{-X}+B_{X\alpha}\left(\sqrt{-X}\right)^2+C_{X\alpha},\label{phi2R0KXs}\\
V(\phi)&=A_{\phi\alpha}\phi^2+B_{\phi\alpha}\phi^4+C_{\phi\alpha},\label{phi2R0Vphis}
\end{align}
\end{subequations}
where all the coefficients are irrelevant to $\beta$.

Furthermore, if we consider the higher-order terms of $\phi$ and $R$ in $p(\phi)$ and $q(R)$, one will have more complicated $K(X)$ and $V(\phi)$.

\section{Localization of the massless graviton}
\label{sec:localization}

From the Schr\"{o}dinger-like equation~\eqref{eq:Schrodinger}, the zero mode of the graviton possesses the following form
\begin{equation}\label{zeromode}
\psi_{0}(z)=N_0\,a(z)^{\frac{3}{2}}f_{R}(z)^{\frac{1}{2}},
\end{equation}
where $N_0$ is the normalization constant.

\begin{figure}[htb]
\centering
\subfigure[\ Ghost-free case]{\label{subfig:fRy'c1}
\includegraphics[width=2.5in]{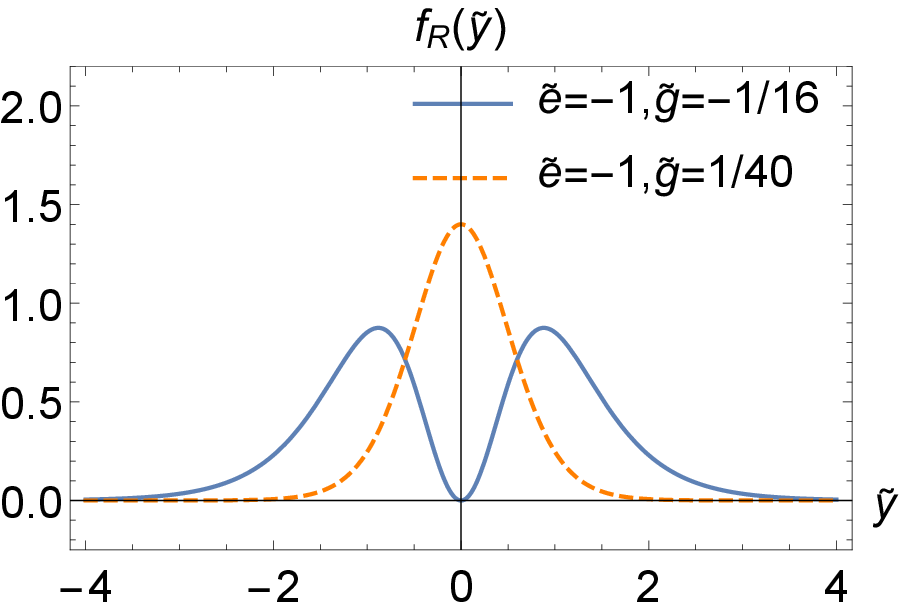}}
\hspace{0.5in}
\subfigure[\ Ghost case]{\label{subfig:fRy'c2}
\includegraphics[width=2.5in]{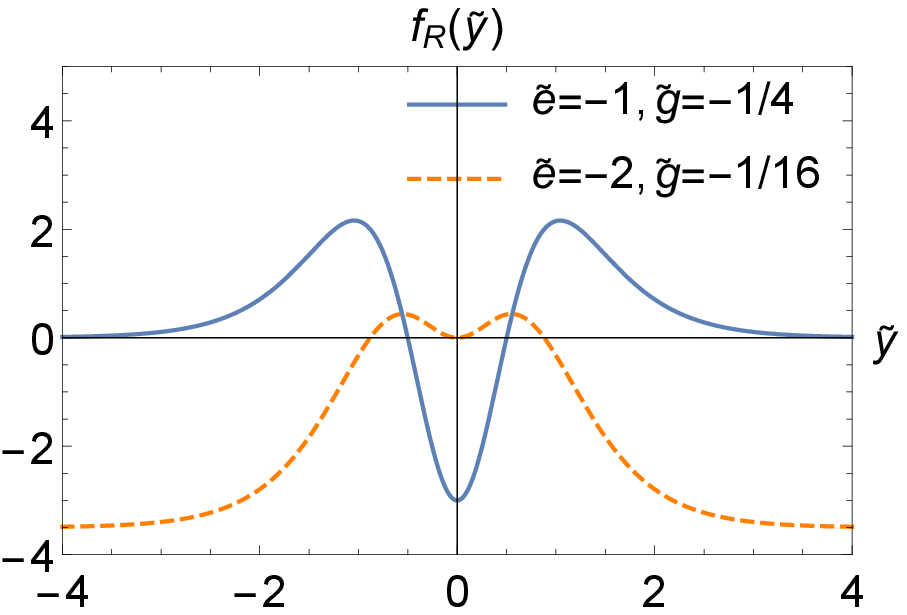}}
\caption{Plots of $f_{R}(\tilde y)$ in Eq.~\eqref{fRys} with $n=1$. (a) The color lines are responsible for the critical lines where $f_{R}$ is always positive even at $\tilde y=0$. These two lines exhibit two ghost-free cases as $f_{R}(\tilde y)>0$. (b) The two color lines illustrate that ghosts are typically contained for $f_{R}(\tilde y)<0$ in some domains of $\tilde y$.}\label{fig:fRy'}
\end{figure}

It should be noted that the condition $f_{R}>0$ must be satisfied in order to avoid the presence of ghost fields. Therefore, the range of values of the parameters $\alpha$, $\beta$, $k$, $v$, and $n$ should be suppressed by the condition $f_{R}>0$. Generally, supposing that the warp factor meets the relation $a(y)=\cosh^{-n}{(ky)}$, and taking notice of the scalar field in Eq.~\eqref{kink} and the particular form of $f(R,\phi,X)$ in Eq.~\eqref{phi2R2}, $f_{R}(y)$ has the following form
\begin{equation}\label{fRy}
f_{R}(y)=\left[\alpha v^2\tanh^2(ky)+1\right]\left[16\beta k^2 n-8\beta k^2n(5n+2)\tanh^2(ky)+1\right].
\end{equation}
Here, we introduce the dimensionless parameters $\tilde y=ky$, $\tilde e=\alpha v^2$, and $\tilde g=k^2\beta$, then Eq.~\eqref{fRy} turns into
\begin{equation}\label{fRys}
f_{R}(\tilde y)=\left[\tilde e\tanh^2(\tilde y)+1\right]\left[16\tilde gn-
8\tilde gn(5n+2)\tanh^2(\tilde y)+1\right].
\end{equation}
From Eq.~\eqref{fRys}, one may prove that the conditions $\tilde e\geqslant-1$ and $-\frac{1}{16n}\leqslant\tilde g\leqslant\frac{1}{40n^2}$ guarantee that it is ghost-free, as shown in Fig.~\ref{fig:fRy'}. Two critical lines are plotted in Fig.~\ref{subfig:fRy'c1}, which are corresponding to the two sets of parameters $\tilde{e}=-1,~\tilde{g}=-1/16$, and $\tilde{e}=-1,~\tilde{g}=1/40$. In fact, if $\tilde{e}$ and $\tilde{g}$ exceed the domain of $\tilde e\geqslant-1$ and $-\frac{1}{16n}\leqslant\tilde g\leqslant\frac{1}{40n^2}$, the system inevitably arises ghosts due to the violation of positivity of $f_{R}$, and two examples for the ghost are shown in Fig.~\ref{subfig:fRy'c2}.

The zero mode of the graviton in Eq.~\eqref{zeromode} takes
\begin{align}\label{zeromodey}
\psi_{0}(z(y))&=N_0\cosh^{-\frac{3}{2}n}(ky)\left[\alpha v^2\tanh^2(ky)+1\right]^\frac{1}{2}
\left[16\beta k^2n-8\beta k^2 n(5n+2)\tanh^2(ky)+1\right]^\frac{1}{2}.
\end{align}
In order to localize the zero mode on the brane, $\psi_{0}(z(y))$ should satisfy the normalization condition
\begin{align}\label{Integration}
&\int_{-\infty}^{+\infty}\psi_0(z){}^2\,\mathrm{d}z=\int_{-\infty}^{+\infty}\psi_0(y){}^2 a(y)^{-1}\,\mathrm{d}y \nonumber \\
=&\int_{-\infty}^{+\infty}\text{sech}^{2n}(ky)\left[\alpha v^2\tanh^2(ky)+1\right]
\left[16\beta k^2n-8\beta k^2n(5n+2)\tanh^2(ky)+1\right]\,\mathrm{d}y<\infty,
\end{align}
which is indeed finite when $n>0$. In other words, the normalized zero mode can be achieved for $n>0$. Hence, the observable four-dimensional gravity is recovered on the brane.

Consequently, we conclude that the zero mode of the graviton is localized on the brane if $n>0$ in the case of $f(R,\phi,X)=(1+\alpha\phi^2)(R+\beta R^2)+K(X)-V(\phi)$. Furthermore, one may verify that the zero mode of the graviton is localized on the brane if $n>0$ for $f(R,\phi,X)=p(\phi)q(R)+K(X)-V(\phi)$ with the finite terms of expansion of $p(\phi)$ and $q(R)$ in Eqs.~\eqref{pphiqR}. In addition, it is worth noting that $f_{R}>0$ is necessary in order to avoid the presence of ghost fields for the arbitrary form of $f(R,\phi,X)$.

\section{Massive KK modes of the graviton}
\label{sec:potential}

\begin{figure}[htb]
\centering
\subfigure[\ $W(\tilde y)/k^2, \psi_0(\tilde y)/\sqrt{k}$]{\label{subfig:Wy}
\includegraphics[width=2.5in]{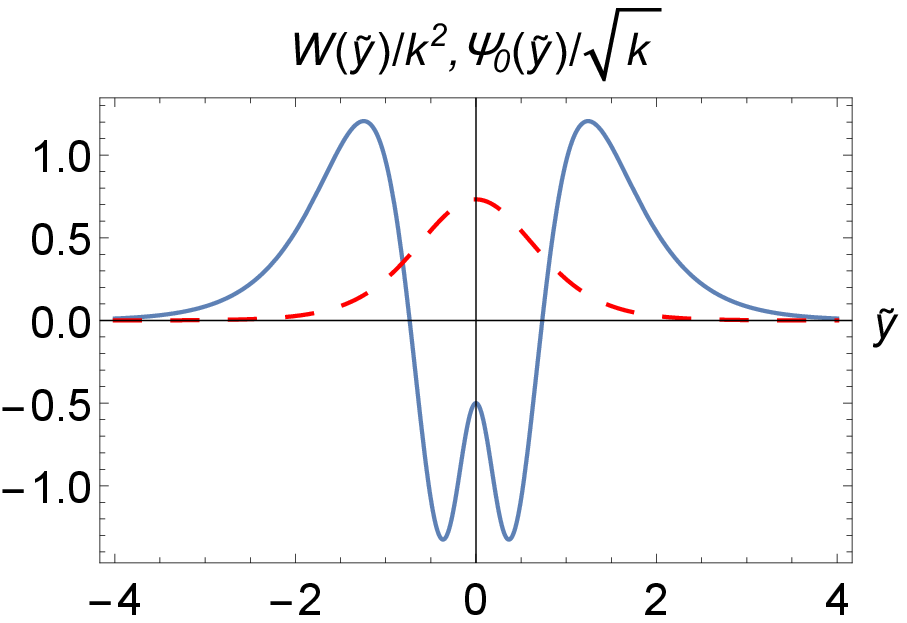}}
\hspace{0.5in}
\subfigure[\ $W(\tilde z)/k^2, \psi_0(\tilde z)/\sqrt{k}$]{\label{subfig:Wz}
\includegraphics[width=2.5in]{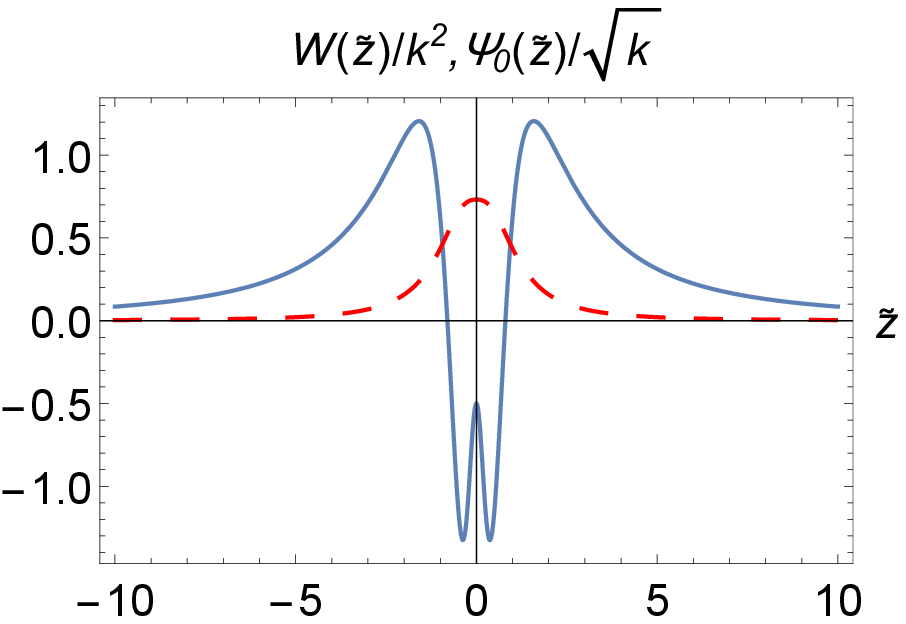}}
\caption{Plots of $W(\tilde y)/k^2$ and $W(\tilde z)/k^2$ (the blue solid lines), and plots of $\psi_0(\tilde y)/\sqrt{k}$ and $\psi_0(\tilde z)/\sqrt{k}$ (the red long-dashed lines), where $W(\tilde y)$ and $W(\tilde z)$ are the effective potentials in Eqs.~\eqref{epotentialy} and \eqref{epotential}, and $\psi_0(\tilde y)$ and $\psi_0(\tilde z)$ are the zero mode of the graviton in Eqs.~\eqref{zeromodey} and \eqref{zeromode}, respectively. Here $\tilde e=2$, and $\tilde g=\frac{1}{40}$.}\label{fig:W}
\end{figure}

In this section, we discuss the effects of the graviton KK modes for the form of $f(R,\phi,X)$ in Eq.~\eqref{phi2R2}. The effective potential~\eqref{epotential} can be transformed as
\begin{equation}\label{epotentialy}
W(z(y))=\frac94\left(\partial_ya\right)^2+\frac32a\partial_{y,y}a+2a\partial_ya\frac{\partial_yf_{R}}{f_{R}}+
\frac12a^2\frac{\partial_{y,y}f_{R}}{f_{R}}-\frac14a^2\frac{(\partial_yf_{R})^2}{f_{R}^2}
\end{equation}
in the coordinate $y$. The effective potentials in Eqs.~\eqref{epotentialy} and \eqref{epotential} and the zero mode of the graviton in Eqs.~\eqref{zeromode} and \eqref{zeromodey} are depicted in Fig.~\ref{fig:W}, for instance taking $a(y)=\cosh^{-1}{(ky)}$ (or $a(z)=\frac{1}{\sqrt{k^2z^2+1}}$ in the $z$ coordinate). As shown in Fig.~\ref{fig:W}, the effective potentials allow a localized zero mode $\psi_0(\tilde{z})$ which is responsible for the four-dimensional Newtonian potential, and a series of continuous massive KK modes $\psi_m(\tilde{z})$ which correct the Newtonian potential.

\begin{figure}[htb]
\centering
\subfigure[\ $\tilde g=0.02$]{\label{subfig:Wy'1}
\includegraphics[width=2.5in]{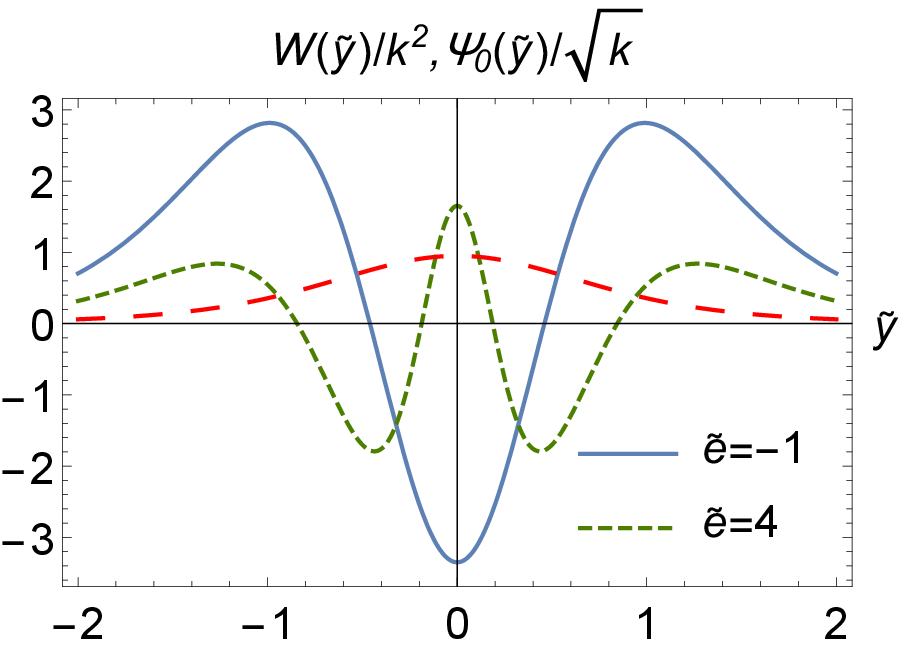}}
\hspace{0.5in}
\subfigure[\ $\tilde g=-0.01$]{\label{subfig:Wy'2}
\includegraphics[width=2.5in]{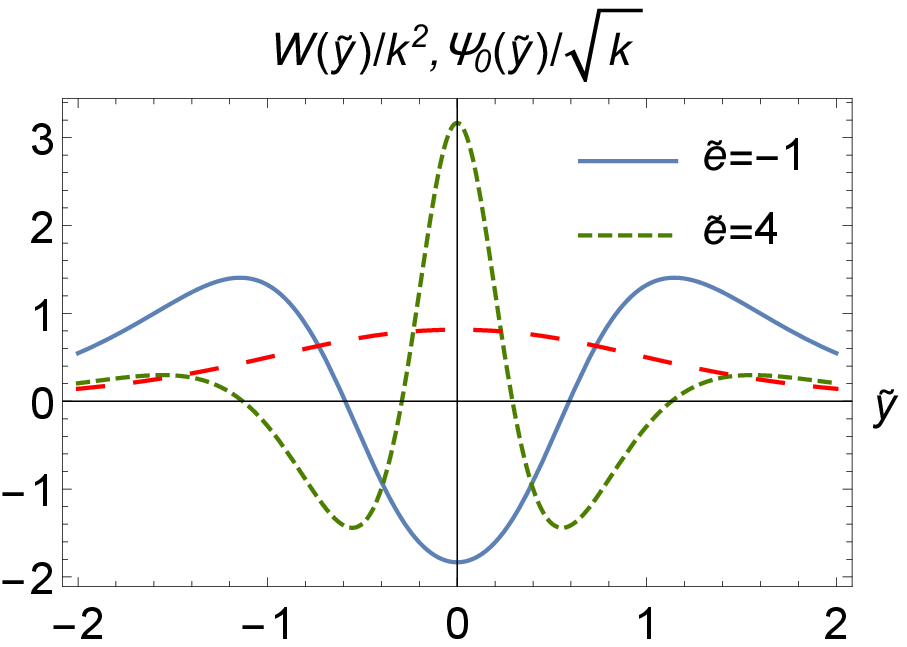}}
\hspace{1in}
\subfigure[\ $\tilde g=-0.03$]{\label{subfig:Wy'3}
\includegraphics[width=2.5in]{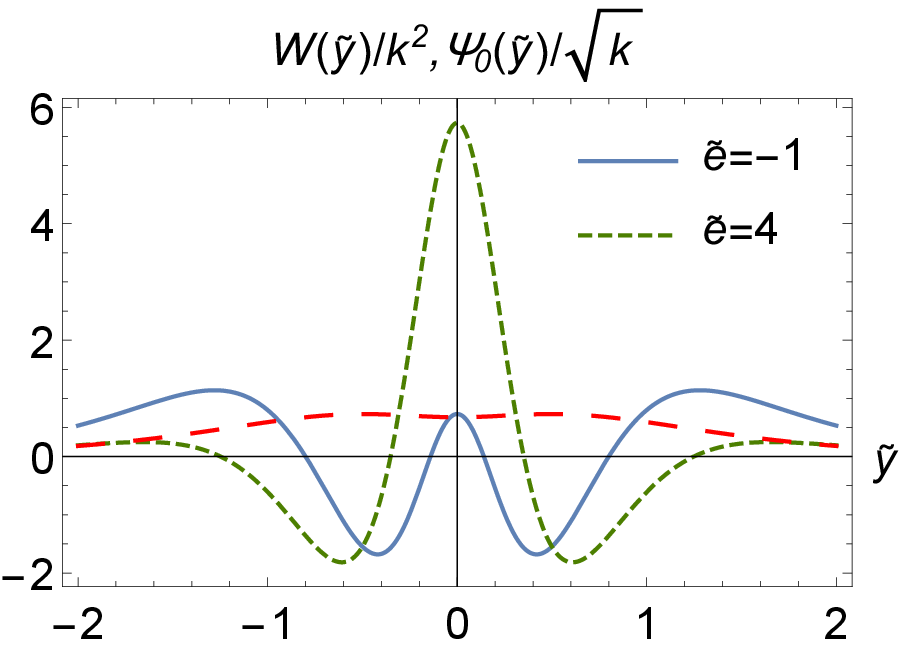}}
\hspace{0.5in}
\subfigure[\ $\tilde g=-0.05$]{\label{subfig:Wy'4}
\includegraphics[width=2.5in]{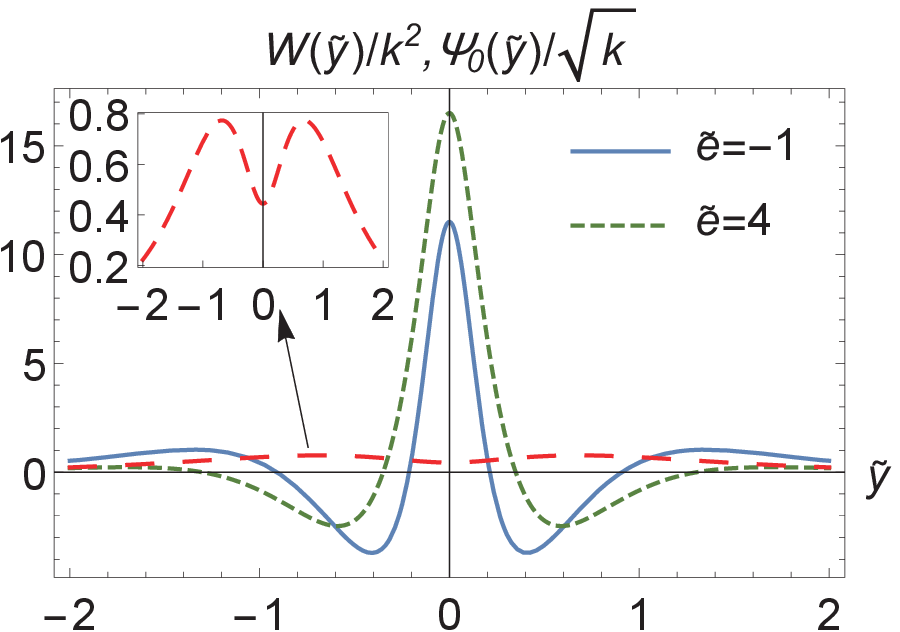}}
\caption{Plots of $W(\tilde y)/k^2$ and $\psi_0(\tilde y)/\sqrt k$ with $n=1$, where $W(\tilde y)$ is the effective potential in Eq.~\eqref{epotentialy} and $\psi_0(\tilde y)$ is the zero mode of the graviton in Eq.~\eqref{zeromodey}. The blue solid lines and green dashed lines display variations. The red long-dashed lines refer to the zero mode of the graviton corresponding to the effective potentials with the blue lines.}\label{fig:Wy'}
\end{figure}

Comparing to the correction of the Newtonian potential in the $f(R)=R+\beta R^2$ case~\cite{Liu:2011lzzl}, we plot the effective potential $W(\tilde y)/k^2$ in Fig.~\ref{fig:Wy'}, from which we find that the potential well will gradually split into two wells and a potential barrier located near $y=0$ will appear with the increasing of $\tilde e$ or the decreasing of $\tilde g$ (for small $\tilde e$). Furthermore, the mass gap depends on the asymptotic behaviour of the effective potential at infinity. the effective potential decays as $1/z^2$ at infinity, and approaches to $0$. So any massive states would be excited states, which go like plane waves, and there is no mass gap. The correction to the Newtonian potential for the $f(R,\phi,X)=(1+\alpha\phi^2)(R+\beta R^2)+K(X)-V(\phi)$ case with small $\tilde e$ around zero is similar to the $f(R)=R+\beta R^2$ case, i.e., the correction $\Delta U(r)\sim1/r^3$~\cite{Liu:2011lzzl}.

\section{Discussion and conclusions}
\label{sec:conclusions}

In summary, we have studied the linear stability of tensor perturbations and the localization of the zero mode of the graviton for general $f(R,\phi,X)$ thick branes, and have found the forms of the function $f(R,\phi,X)$ supporting domain wall configurations. The equation of the TT tensor modes was obtained for $f(R,\phi,X)$ thick branes, whose extra-dimensional part can be converted to a Schr\"{o}dinger-like equation. The ``Hamiltonian" in the Schr\"{o}dinger-like equation was factorized into a supersymmetric form which results in no the gravitational tachyon mode with normalizable negative energy, and therefore the brane systems are stable against the tensor perturbations under the condition $f_{R}>0$. This conclusion indicates that the linear stability of the tensor perturbations for $f(R,\phi,X)$ thick branes can be applied to a wide range of thick brane models such as $f(R)$ thick branes, scalar-tensor thick branes, and $K$-field thick branes.

In order to obtain domain wall configurations, we found a typical form of $f(R,\phi,X)=(1+\alpha\phi^2)(R+\beta R^2)+K(X)-V(\phi)$ by performing the reconstruction technique that one can find the original action by substituting the solution of the warp factor and scalar field into the general equations of motion. Clearly, this form can degenerate into $f(R)$, scalar-tensor, and GR with the non-canonical scalar field. These results imply that non-canonical scalar fields are of significance for thick branes. As shown previously, the reconstruction technique is an alternative approach to explore new brane configurations.

For the case of $f(R,\phi,X)=(1+\alpha\phi^2)(R+\beta R^2)+K(X)-V(\phi)$ and $a(y)=\cosh^{-n}{(ky)}$, the graviton zero mode can be localized on the brane for the condition $n>0$. We obtained the conditions which avoid the presence of ghosts. In fact, the zero mode of the graviton is invariably localized if $n>0$ for $f(R,\phi,X)=p(\phi)q(R)+K(X)-V(\phi)$, where there are finite terms in the expansions of $p(\phi)$ and $q(R)$. With regard to the massive KK modes of the graviton, the correction to the Newtonian potential is $\Delta U(r)\sim1/r^3$ for small $\tilde{e}$ around zero.

\section*{Acknowledgements}

This work is supported in part by the National Natural Science Foundation of China (Grants Nos. 11875151, 11522541 and 10905027) and the Fundamental Research Funds for the Central Universities (Grants Nos. lzujbky-2018-k11 and lzujbky-2017-it69). B.-M. Gu. is supported by a scholarship granted by the China Scholarship Council (CSC, No. 201706180072).

\section*{Appendix: a procedure for the solution}
\label{Apendix}

Here, we provide some intermediate steps from Eqs.~\eqref{eq:eq1MN} and \eqref{eq:eq2phi} to the results~\eqref{phi2R2_KV} under assumptions~\eqref{phi2R2}, \eqref{warpfactor}, and \eqref{kink}. From the form of $f(R,\phi,X)$~\eqref{phi2R2}, one may calculate:
\begin{subequations}
\begin{align}
  f =& 4\left(\alpha \phi^2+1\right) \left(\frac{2 a''}{a}+\frac{3 a'^2}{a^2}\right) \left[4 \beta \left(\frac{2 a''}{a}+\frac{3 a'^2}{a^2}\right)- 1\right]+K-V, \\
  f_{R} =& \left(\alpha \phi^2+1\right) \left[1-8 \beta \left(\frac{2 a''}{a}+\frac{3 a'^2}{a^2}\right)\right], \\
  f_{R}' =& 2 \alpha \phi \phi' \left[1-8 \beta \left(\frac{2 a''}{a}+\frac{3 a'^2}{a^2}\right)\right]-8 \beta \left(\alpha \phi^2+1\right) \left(\frac{2 a^{(3)}}{a}-\frac{6 a'^3}{a^3}+\frac{4 a' a''}{a^2}\right), \\
  f_{R}'' =& 2 \alpha (\phi \phi''+\phi'^2) \left[1-8 \beta \left(\frac{2 a''}{a}+\frac{3 a'^2}{a^2}\right)\right]-32 \alpha \beta \phi \phi' \left(\frac{2 a^{(3)}}{a}-\frac{6 a'^3}{a^3}+\frac{4 a' a''}{a^2}\right)\nonumber\\&-8 \beta \left(\alpha  \phi^2+1\right) \left(\frac{2 a^{(4)}}{a}+\frac{4 a''^2}{a^2}+\frac{18 a'^4}{a^4}+\frac{2 a^{(3)} a'}{a^2}-\frac{26 a'^2 a''}{a^3}\right), \\
  f_{X} =& -\frac{K'}{\phi' \phi''}, \\
  f_{X}' =& -\frac{K''}{\phi' \phi''}+\frac{K'}{\phi'^2}+\frac{\phi^{(3)} K'}{\phi' \phi''^2}, \\
  f_{\phi} =& 8 \alpha \phi \left(\frac{2 a''}{a}+\frac{3 a'^2}{a^2}\right) \left[4 \beta \left(\frac{2 a''}{a}+\frac{3 a'^2}{a^2}\right)- 1\right]-\frac{V'}{\phi'},
\end{align}
\end{subequations}
where the superscript $^{(i)}$ ($i=3, 4, \cdots$) denotes the $i$-th order derivatives with respect to $y$. They are functions of $y$ only.
One can obtain the concrete form of Eqs.~\eqref{eq:eq1MN} and \eqref{eq:eq2phi} as
\begin{subequations}\label{eq:eq1MN'}
\begin{gather}
  4\left(\alpha \phi^2+1\right) \left(\frac{2 a''}{a}+\frac{3 a'^2}{a^2}\right)\left[4 \beta \left(\frac{2 a''}{a}+\frac{3 a'^2}{a^2}\right)- 1\right]+2 \left(\alpha \phi^2+1\right) \left(\frac{a''}{a}+\frac{3 a'^2}{a^2}\right) \left[1-8 \beta \left(\frac{2 a''}{a}+\frac{3 a'^2}{a^2}\right)\right]\nonumber\\-\frac{6 a'}{a} \left\{2 \alpha \phi \phi' \left[1-8 \beta  \left(\frac{2 a''}{a}+\frac{3 a'^2}{a^2}\right)\right]-8 \beta \left(\alpha \phi^2+1\right) \left(\frac{2 a^{(3)}}{a}-\frac{6 a'^3}{a^3}+\frac{4 a' a''}{a^2}\right)\right\}\nonumber\\-2 \left\{2 \alpha (\phi \phi''+\phi'^2) \left[1-8 \beta \left(\frac{2 a''}{a}+\frac{3 a'^2}{a^2}\right)\right]-32 \alpha \beta \phi \phi' \left(\frac{2 a^{(3)}}{a}-\frac{6 a'^3}{a^3}+\frac{4 a' a''}{a^2}\right)\right.\nonumber\\ \left.-8 \beta \left(\alpha  \phi^2+1\right) \left(\frac{2 a^{(4)}}{a}+\frac{4 a''^2}{a^2}+\frac{18 a'^4}{a^4}+\frac{2 a^{(3)} a'}{a^2}-\frac{26 a'^2 a''}{a^3}\right)\right\}+K-V=0, \\
  4\left(\alpha \phi^2+1\right) \left(\frac{2 a''}{a}+\frac{3 a'^2}{a^2}\right)\left[4 \beta \left(\frac{2 a''}{a}+\frac{3 a'^2}{a^2}\right)- 1\right]+\frac{8 a''}{a}\left(\alpha \phi^2+1\right) \left[1-8 \beta \left(\frac{2 a''}{a}+\frac{3 a'^2}{a^2}\right)\right]\nonumber\\-\frac{8 a'}{a} \left\{2 \alpha \phi \phi' \left[1-8 \beta \left(\frac{2 a''}{a}+\frac{3 a'^2}{a^2}\right)\right]-8 \beta \left(\alpha \phi ^2+1\right) \left(\frac{2 a^{(3)}}{a}-\frac{6 a'^3}{a^3}+\frac{4 a' a''}{a^2}\right)\right\}+K-V=\frac{K' \phi'}{\phi''},
\end{gather}
\end{subequations}
and
\begin{equation}\label{eq:eq2phi'}
  \left(-\frac{K''}{\phi' \phi''}+\frac{K'}{\phi'^2}+\frac{\phi ^{(3)} K'}{\phi' \phi''^2}\right)\phi'-\frac{K'}{\phi' \phi''}\left(\frac{4 a'\phi'}{a}+\phi''\right)+8 \alpha \phi \left(\frac{2 a''}{a}+\frac{3 a'^2}{a^2}\right) \left[4 \beta \left(\frac{2 a''}{a}+\frac{3 a'^2}{a^2}\right)- 1\right] -\frac{V'}{\phi'}=0,
\end{equation}
by taking the assumption~\eqref{phi2R2} into account. However, only two equations are independent in three non-linear differential equations~\eqref{eq:eq1MN'} and \eqref{eq:eq2phi'} where the highest-order derivatives are fourth-order and second-order respectively with regard to the warp factor and scalar field. Fortunately, we concentrate on $K$ and $V$ rather than the warp factor and scalar field because of employing the reconstruction technique. One finds that the highest-order derivatives are explicitly second-order and zero-order with regard to $K$ and $V$. For the sake of simplicity, we use equations~\eqref{eq:eq1MN'} to work out $K$ and $V$ and use equation~\eqref{eq:eq2phi'} to verify the solutions. Note that we preset the warp factor and scalar field as ~\eqref{warpfactor} and ~\eqref{kink}. We obtain
\begin{subequations}
\begin{align}
  K(y)=& \left[A_K+B_K\cosh (2 k y)+C_K\cosh (4 k y)\right]\text{sech}^6(k y)+D_K,\label{Ky}\\
  V(y)=& A_V \text{sech}^2(k y)+ B_V \text{sech}^4(k y)+ C_V \text{sech}^6(k y)+D_V,\label{Vy}
\end{align}
\end{subequations}
where the coefficients are given by
\begin{align}
  A_K =& -\frac{9k^2 n}{4}-\frac{k^2 v^2 n}{4}\alpha +2 k^4 n^2 (5 n-16)\beta+\frac{2 k^4 v^2 n \left[n(304-25 n)+128\right]}{3} \alpha \beta, \nonumber\\
  B_K =& -3 k^2 n-\frac{k^2 v^2(n+2)}{2}\alpha +4 k^4 n \left(5 n^2+4\right)\beta-8 k^4 v^2 n (31 n+14) \alpha \beta, \nonumber\\
  C_K =& -\frac{3k^2 n}{4}-\frac{k^2 v^2(n+4)}{4}\alpha+2 k^4 n \left[n (5 n+16)+8\right]\beta+2 k^4 v^2 n \left[n (36-5 n)+8\right] \alpha \beta, \nonumber\\
  D_K =& c, \nonumber\\
  A_V =& 12 k^2 v^2 n^2 +8 k^2 n (3 n+2)\alpha -32 k^4 n^2 \left[n (5 n+16)+8\right]\beta \nonumber\\&-16 k^4 v^2 n^2 \left[3 n (5 n+24)+16\right]\alpha \beta, \nonumber\\
  B_V =& -k^2 v^2(3 n+2) (4 n+3)\alpha +8 k^4 n (n+6) (2 n+1) (5 n+2)\beta \nonumber\\&+16 k^4 v^2 n \left\{n \left[n (15 n+139)+111\right]+22\right\}\alpha \beta, \nonumber\\
  C_V =& -\frac{16 k^4 v^2 n (5 n+2) \left[n (3 n+40)+40\right]}{3}\alpha \beta, \nonumber\\
  D_V =& c-12 k^2 n^2-12 k^2 v^2 n^2 \alpha+80 k^4 n^4\beta +80 k^4 v^2 n^4 \alpha \beta. \nonumber
\end{align}
Here $\alpha$, $\beta$, $k$, $v$, and $c$ are real parameters. This solution exactly satisfies Eq.~\eqref{eq:eq2phi'}. According to $X=-\phi(y)'^2/2$ and $\phi(y)=v\tanh(ky)$, the inverse functions (a branch) would be
\begin{equation}\label{invX}
  y=\frac{1}{k}\cosh ^{-1}\left(\sqrt[4]{-\frac{1}{2 X}} \sqrt{k v}\right),
\end{equation}
and
\begin{equation}\label{invphi}
  y=\frac{1}{k}\tanh ^{-1}\left(\frac{\phi }{v}\right).
\end{equation}
Substituting of Eqs.~\eqref{invX} and \eqref{invphi} into $~\eqref{Ky}$ and $~\eqref{Vy}$ respectively, one can find
\begin{subequations}
\begin{align}
K(X)&=A_X\sqrt{-X}+B_X\left(\sqrt{-X}\right)^2+C_X\left(\sqrt{-X}\right)^3+D_X, \\
V(\phi)&=A_{\phi}\phi^2+B_{\phi}\phi^4+C_{\phi}\phi^6+D_{\phi},
\end{align}
\end{subequations}
where the coefficients are given by
\begin{align}
A_X&=-\frac{6\sqrt{2}kn}{v}-2\sqrt{2}kv\left(n+4\right)\alpha+\frac{16\sqrt{2}k^3n\left(5n^2+16n+8\right)}{v}\beta
\nonumber \\ &\quad+16\sqrt{2}k^3vn\left(-5n^2+36n+8\right)\alpha\beta, \nonumber \\
B_X&=2\left(n+6\right)\alpha-\frac{16k^2n\left(n+6\right)\left(5n+2\right)}{v^2}\beta+
32k^2n\left[n\left(5n-67\right)-22\right]\alpha\beta, \nonumber \\
C_X&=-\frac{32\sqrt{2}kn\left(n-20\right)\left(5n+2\right)}{3v}\alpha\beta, \nonumber \qquad\qquad\qquad D_X=c,
\end{align}
and
\begin{align}
A_{\phi}&=-\frac{12k^2n^2}{v^2}+6k^2\left(3n+2\right)\alpha-\frac{16k^4n\left[n\left(37n+40\right)+12\right]}{v^2}\beta
\nonumber \\ &\quad+32k^4n\left(37n+18\right)\alpha\beta, \nonumber \\
B_{\phi}&=-\frac{k^2\left(3n+2\right)\left(4n+3\right)}{v^2}\alpha+
\frac{8k^4n\left(n+6\right)\left(2n+1\right)\left(5n+2\right)}{v^4}\beta \nonumber \\
&\quad-\frac{16k^4n\left[n\left(67n+169\right)+58\right]}{v^2}\alpha\beta, \nonumber \\
C_{\phi}&=\frac{16k^4n\left(5n+2\right)\left[n\left(3n+40\right)+40\right]}{3v^4}\alpha\beta, \nonumber \\
D_{\phi}&=c-k^2v^2\left(n+6\right)\alpha+8k^4n\left[n\left(5n+24\right)+12\right]\beta-
\frac{16k^4v^2n\left[5\left(n-1\right)n+14\right]}{3}\alpha\beta. \nonumber
\end{align}

\end{document}